\documentclass[12pt]{article}
\usepackage{fullpage}
\usepackage{cite}
\usepackage{amsmath}
\usepackage{amssymb}
\usepackage{epsfig}
\title{}
\date{}
\def\para{\\ [-2mm]}

\def\mands{ (k_1+k_2)^2 }
\def\mandt{ (k_1+k_4)^2 }
\def\eps  {\epsilon}
\def\cN{ {\cal N} }
\def\Epsilon{  {\cal E} }
\def\epseps#1#2{ \epsilon_{#1} \cdot \epsilon_{#2} }
\def\epsk#1#2{ \epsilon_{#1} \cdot k_{#2} }
\def\kk#1#2{ k_{#1} \cdot k_{#2} }
\def\KK#1#2{ (k_{#1} \cdot k_{#2} - \kappa_{#1} \kappa_{#2} )}

\def\be{\begin{equation}}
\def\ee{\end{equation}}
\def\ba{\begin{eqnarray}}
\def\ea{\end{eqnarray}}
\def\nl{\nonumber\\}
\def\nn{\nonumber}
\def\eqn#1{eq.~(\ref{#1})} 
\def\eqns#1#2{eqs.~(\ref{#1}) and~(\ref{#2})}

\def\id{  {\mathsf{1}\kern -3pt \mathsf{l} } }
\def\half{  {1\over 2} }

\def\Tr {\mathop{\rm Tr}\nolimits}

\begin{document}
\bibliographystyle{nb}

\titlepage
\begin{flushright}
BOW-PH-161\\
\end{flushright}

\vspace{3mm}

\begin{center}

{\Large\bf\sf
Amplitudes for massive vector and scalar bosons
\\[1mm]
in spontaneously-broken gauge theory 
\\[3mm]
from the CHY representation
}

\vskip 1.5cm

{\sc Stephen G. Naculich }

\vskip 3mm

{\it
Department of Physics\\
Bowdoin College\\
Brunswick, ME 04011, USA
}

\vskip 3mm 

{\tt naculich@bowdoin.edu }

\end{center}

\vskip 1.5cm

\begin{abstract}
In the formulation of Cachazo, He, and Yuan, tree-level amplitudes for 
massless particles in gauge theory and gravity can be expressed as 
rational functions of the Lorentz invariants 
$k_a \cdot k_b$,  $\epsilon_a \cdot k_b$,  and $\epsilon_a \cdot \epsilon_b$,  
valid in any number of spacetime dimensions.
We use dimensional reduction of higher-dimensional amplitudes of particles 
with internal momentum $\kappa$ to obtain amplitudes for massive particles 
in lower dimensions.  In the case of gauge theory, we argue that these 
massive amplitudes belong to a theory in which the gauge symmetry is 
spontaneously broken by an adjoint Higgs field.  Consequently, we show that 
tree-level $n$-point amplitudes containing massive vector and scalar bosons
in this theory can be obtained by simply replacing $k_a \cdot k_b$ with 
$k_a \cdot k_b - \kappa_a \kappa_b $ in the corresponding massless amplitudes,
where the masses of the particles are given by $|\kappa_a|$.
\end{abstract}

\vspace*{0.5cm}

\vfil\break

In a series of papers over the last few years,
Cachazo, He, and Yuan (CHY) have developed a new representation
for the tree-level amplitudes of massless particles in a variety of theories, 
including Yang--Mills theory and Einstein 
gravity \cite{Cachazo:2013hca,Cachazo:2013iea,Cachazo:2014nsa},
as well as the nonlinear $\sigma$ model, the Dirac--Born--Infeld theory, 
and more exotic models \cite{Cachazo:2014xea}.
In this approach, the ingredients for an $n$-point amplitude 
are the positions $\sigma_a$ of $n$ punctures on a sphere,
as well as Lorentz invariants 
of the momenta $k^\mu_a$ and polarizations $\eps^\mu_a$ of the particles.
(The graviton polarization is expressed as 
$\eps^{\mu\nu}_a = \eps^\mu_a \tilde{\eps}^\nu_a$.)
The CHY representation involves an integral 
over the moduli space of the punctured sphere, which can be 
evaluated \cite{Cachazo:2013iea,Cachazo:2014xea,Kalousios:2015fya,Cachazo:2015nwa}
to yield a rational expression of $\kk{a}{b}$, $\epsk{a}{b}$, and $\epseps{a}{b}$
for the $n$-point tree-level amplitude.
\para

A notable feature of this formulation is its 
independence of the number of spacetime dimensions.
One can therefore dimensionally reduce CHY representations 
of gravity or Yang--Mills amplitudes in $(d+M)$ dimensions
to obtain mixed amplitudes in $d$ dimensions
containing gravitons, gauge bosons, and massless scalars 
by choosing the $(d+M)$-dimensional polarizations $\Epsilon_a$
to lie in either the $d$- or $M$-dimensional 
subspace \cite{Cachazo:2014nsa,Cachazo:2014xea}.
\para

Dimensional reduction can also be used to endow particles with mass,
and therefore to obtain CHY representations for massive particles 
in $d$ dimensions \cite{Naculich:2015zha}.
(Previous extensions of the CHY approach to massive particles were 
considered in refs.~\cite{Dolan:2013isa,Dolan:2014ega,Naculich:2014naa}.)
The momentum of a particle in $(d+M)$ dimensions can be written 
\be
K_a ~=~ (k_a | \kappa_a)
\ee
where $k_a$ and $\kappa_a$ are the components of momentum 
in $d$- and $M$-dimensional subspaces respectively;
we will refer to $\kappa_a$ as the {\it internal momentum} of a particle,
and regard it as a fixed quantity.
A massless particle in $(d+M)$ dimensions with momentum $K_a$
corresponds to a particle in $d$ dimensions with mass $m_a= |\kappa_a|$.
This approach is similar to that used 
in refs.~\cite{Alday:2009zm,Henn:2010bk,Henn:2010ir},
where a fifth dimension was introduced
as a massive infrared regulator for planar loop-level $\cN=4$ amplitudes.
See also, for example, 
refs.~\cite{Selivanov:1999ie,Boels:2010mj,Bern:2010qa,Bjerrum-Bohr:2013bxa,Plefka:2014fta},
in which massive particles are represented as higher-dimensional massless particles.
\para

The amplitude for massless gauge bosons in $(d+M)$ dimensions,
which is a rational function of $K_a \cdot K_b$, 
$\Epsilon_a \cdot K_b$, and $\Epsilon_a \cdot \Epsilon_b$,
can be used to obtain an expression for 
the amplitude for massive gauge bosons in $d$ dimensions
by choosing $\Epsilon_a = (\eps_a|0)$, 
so that\footnote{The relative minus sign in $K_a \cdot K_b$
arises because we use a mostly-minus metric for $K_a$ and $k_a$, 
but an all-plus metric for the internal components $\kappa_a$.} 
\be
K_a \cdot K_b = \kk{a}{b} - \kappa_a \cdot \kappa_b , \qquad\qquad
\Epsilon_a \cdot K_b  = \epsk{a}{b}, \qquad\qquad
\Epsilon_a \cdot \Epsilon_b  = \epseps{a}{b}  \,.
\label{massivevector}
\ee
In other words, we simply replace 
$\kk{a}{b}$ with $ \kk{a}{b} - \kappa_a \cdot \kappa_b$
in the $d$-dimensional massless gauge boson amplitude.
The question remains: to what theory do these massive amplitudes belong?
\para

Not all $d$-dimensional amplitudes of massive particles can be obtained 
from dimensional reduction of $(d+M)$-dimensional massless amplitudes, 
due to the constraints arising from internal momentum conservation.
Overall momentum conservation demands  $\sum_{a=1}^n \kappa_a = 0$, 
placing a restriction on the masses of a dimensionally-reduced $n$-particle 
amplitude.  Furthermore, internal momentum conservation must be satisfied
at each vertex of the tree diagrams that contribute to the amplitude,
constraining the mass of any particle propagating in intermediate channels.
For example, if particles $a$ and $b$ couple into an intermediate channel,
the mass of the particle in that channel must be given by $|\kappa_a + \kappa_{b}|$.
In ref.~\cite{Naculich:2015zha},
it was shown that this constraint is automatically satisfied
in an amplitude with no more than three massive particles,
in which the remaining massless particles are flavor-preserving,
and therefore such amplitudes can be given a CHY representation.
But for more general massive amplitudes (e.g. in the standard model), 
the internal momentum constraints cannot be satisfied.
\para 

It is the purpose of this note to observe that the internal momentum 
constraints {\it can} be satisfied in a gauge theory in which the 
$U(N)$ symmetry is spontaneously broken by an adjoint Higgs field $H$.
If the Higgs field has vacuum expectation value 
$\langle H \rangle = {\rm diag}(v_1, v_2, \cdots, v_N)$,
the $U(N)$ symmetry is broken to $U(1)^N$,
and the off-diagonal gauge bosons $W_i^{~j}$ obtain masses  $g |v_i - v_j|$
from the $\Tr ( D_\mu H)^2 $ term in the Lagrangian.
If some of the $v_i$ are degenerate, then a larger symmetry will be left unbroken.
(This is essentially the theory considered in
refs.~\cite{Alday:2007hr,Kawai:2007eg,Schabinger:2008ah,McGreevy:2008zy,Berkovits:2008ic,Alday:2009zm,Boels:2010mj,Plefka:2014fta},
where it is embedded in an $\cN=4$ supersymmetric theory on the Coulomb branch.
In the string theory picture, the $v_i$ correspond to the positions of D3 branes
in the fifth dimension, and massive gauge bosons correspond to 
strings extending between separated D3 branes.
See also refs.~\cite{Craig:2011ws,Kiermaier:2011cr} 
for tree-level massive amplitudes on the Coulomb branch of this theory.)
\para

At a triple-gauge-boson vertex, $W_i^{~j} $ and $W_j^{~k}$ couple to $W_i^{~k}$.
If we identify the internal momentum\footnote{now restricted to one dimension}
$\kappa_{ij}$ of $W_i^{~j}$ with $ g(v_i - v_j)$,
then internal momentum conservation $\kappa_{ij}+\kappa_{jk} = \kappa_{ik}$ 
automatically holds at each vertex
(and similarly at four-gauge-boson vertices).
This suggests that dimensionally-reduced amplitudes with internal momenta $\kappa$
are equivalent to amplitudes of massive gauge bosons in this 
spontaneously-broken theory.
Moreover, this implies that massive amplitudes in this theory
can be obtained from the corresponding massless amplitudes of the
unbroken theory, 
expressed in terms of 
$\kk{a}{b}$, $\epsk{a}{b}$, and $\epseps{a}{b}$,
by simply replacing 
$k_a \cdot k_b$ with $k_a \cdot k_b - \kappa_a \kappa_b $.
To illustrate this in a simple case, we evaluate the 
amplitude of four massive gauge bosons
in the spontaneously-broken theory 
\begin{align}
&\langle  
W_{i_1}^{~i_2} (k_1,\eps_1) 
W_{i_2}^{~i_3} (k_2,\eps_2)  
W_{i_3}^{~i_4} (k_3,\eps_3) 
W_{i_4}^{~i_1} (k_4,\eps_4) 
\rangle
\nn\\
&~~~~~~~~~~~= g^2 \left[ {N_s \over \mands - m_s^2}
+ {N_t \over \mandt  - m_t^2}
+ ({\rm contact~term}) \right]
\end{align}
which have contributions from $s$- and $t$-channel exchange
as well as a four-gauge-boson contact term.
For brevity, we include below only those terms in the numerators that depend
on $\kk{a}{b}$;  the omitted terms are the same as those in the unbroken theory:
\begin{align}
N_s  &=  {(\eps_1 \cdot \eps_2)  \ ( \eps_3 \cdot \eps_4) \   \over 2}
\left[ ( k_2 -k_1) \cdot (k_4 - k_3)
      + {(m_2^2-m_1^2) (m_4^2 - m_3^2) \over m_s^2 }
\right] ~+~ \cdots \,,
\nn\\
N_t &=  {(\eps_1 \cdot \eps_4)  \ ( \eps_2 \cdot \eps_3) \   \over 2}
\left[  (k_2 -k_3) \cdot (k_4- k_1)
            + { (m_1^2-m_4^2) (m_3^2 - m_2^2) \over m_t^2 }
\right]   ~+~ \cdots \,.
\end{align}
Here $m_a= |\kappa_a|$ with $\kappa_a \equiv g( v_{i_a} - v_{i_{a+1}})$,
where $i_5 \equiv i_1$ so that $\sum_{a=1}^4 \kappa_a = 0$.
Up to this point, we have not made any assumptions 
about the masses of the $s$ and $t$-channel intermediate particles.
Since the particles propagating in those channels are
$W_{i_1}^{~i_3}$, and $W_{i_2}^{~i_4}$ respectively,
we have
\be 
m_s^2 = g^2 (v_{i_1} - v_{i_3} )^2 = (\kappa_1 + \kappa_2)^2, \qquad\qquad
m_t^2 = g^2 (v_{i_2} - v_{i_4} )^2 = (\kappa_2 + \kappa_3)^2
\ee
and the numerators above simplify to 
\begin{align}
&N_s  =  \half (\eps_1 \cdot \eps_2)  \ ( \eps_3 \cdot \eps_4) \  
\Bigl[ ( k_2 -k_1) \cdot (k_4 - k_3)
      - (\kappa_2 -\kappa_1) (\kappa_4 - \kappa_3)
\Bigr] ~+~ \cdots \,,
\nn\\
&N_t =   \half(\eps_1 \cdot \eps_4)  \ ( \eps_2 \cdot \eps_3) \  
\Bigl[  (k_2 -k_3) \cdot (k_4- k_1)
    -        (\kappa_2 -\kappa_3) (\kappa_4- \kappa_1)
\Bigr]  ~+~ \cdots \,.
\end{align}
We assemble all the pieces and use 
\begin{align}
\mands -m_s^2 &=  2 \KK{1}{2} =  2 \KK{3}{4}\,, \nn\\[2mm]
\mandt -m_t^2 &=  2 \KK{1}{4} = 2 \KK{2}{3} 
\end{align}
to obtain 
\be
\langle  
W_{i_1}^{~i_2} (k_1,\eps_1) 
W_{i_2}^{~i_3} (k_2,\eps_2)  
W_{i_3}^{~i_4} (k_3,\eps_3) 
W_{i_4}^{~i_1} (k_4,\eps_4) 
\rangle
= -  g^2 { {\bf K} \over  \KK{2}{3} \KK{3}{4}  }
\label{fourpoint}
\ee
with
\ba
{\bf K}  &=-& 
 \Big\{  
     \KK{1}{3} \KK{2}{3}  ~\epseps{1}{ 2} ~\epseps{3}{ 4} \nl
&& + \KK{2}{3} \KK{3}{4}  ~\epseps{1}{ 3} ~\epseps{2}{ 4} \nl
&&+  \KK{1}{3} \KK{3}{4}  ~\epseps{1}{ 4} ~\epseps{2}{ 3} \nl
&& +\Bigl[\KK{1}{ 3} ~\epsk{1}{ 4}~ \epsk{2}{ 3} +\KK{2}{ 3} ~\epsk{1}{ 3} ~\epsk{2}{ 4}  \Bigr]~\epseps{3}{ 4}  \nl
&& +\Bigl[\KK{2}{ 3}~ \epsk{1}{ 2} ~\epsk{3}{ 4} + \KK{3}{ 4} ~\epsk{1}{ 4} ~\epsk{3}{ 2} \Bigr]~\epseps{2}{ 4}  \nl
&& +\Bigl[\KK{1}{ 3} ~ \epsk{1}{ 2}~ \epsk{4}{ 3} + \KK{3}{ 4}~ \epsk{1}{ 3}~ \epsk{4}{ 2}\Bigr]~\epseps{2}{ 3}  \nl
&& +\Bigl[ \KK{1}{ 3}~ \epsk{2}{ 1}~ \epsk{3}{ 4} +\KK{3}{ 4}~ \epsk{2}{ 4} ~\epsk{3}{ 1} \Bigr]~\epseps{1}{ 4}   \nl
&& +\Bigl[ \KK{2}{ 3}~ \epsk{2}{ 1}~ \epsk{4}{ 3}+ \KK{3}{ 4}~ \epsk{2}{ 3} ~\epsk{4}{ 1} \Bigr] ~\epseps{1}{ 3}  \nl
&& +\Bigl[ \KK{1}{ 3}~ \epsk{3}{ 2}~ \epsk{4}{ 1} + \KK{2}{ 3}~ \epsk{3}{ 1}~ \epsk{4}{ 2}\Bigr]~\epseps{1}{ 2}   
\Big\}  \,.
\label{K}
\ea
As expected, this is simply the four-gluon amplitude with 
$\kk{a}{b}$ replaced by $ \kk{a}{b} - \kappa_a \kappa_b$.
\para

Mixed amplitudes containing both massive vector and adjoint
scalar bosons in $d$ dimensions can also be obtained via 
dimensional reduction of 
massless gauge amplitudes in $d+M$ dimensions
\cite{Cachazo:2014xea,Naculich:2015zha}.
Massive adjoint scalar bosons arise from particles 
whose $(d+M)$-dimensional polarization and momentum vectors
are given by\footnote{We choose the internal polarization of the 
would-be scalars to be orthogonal to the internal 
momenta of all the particles.}
\be
\Epsilon_a = (0 | 0, e_a) ,\qquad\qquad 
K_a ~=~ (k_a | \kappa_a,0).
\ee
Thus, invariants involving the adjoint scalars are given by
\be
K_a \cdot K_b = \kk{a}{b} - \kappa_a \cdot \kappa_b, \qquad\qquad
\Epsilon_a \cdot K_b  = 0, \qquad\qquad
\Epsilon_a \cdot \Epsilon_b  = e_a \cdot e_b  \,.
\ee
Again this implies that the amplitudes for massive vectors and scalars
can be obtained from the corresponding massless amplitudes 
by simply replacing $\kk{a}{b}$ with $ \kk{a}{b} - \kappa_a \cdot \kappa_b$.
\para

The field theory corresponding to these dimensionally-reduced 
amplitudes contains adjoint scalar fields $\Phi_i^{~j}$,
whose masses $g |v_i - v_j|$ arise from the 
$ g^2 \Tr \left( [\Phi, H] ^2 \right) $ term in the 
Lagrangian \cite{Alday:2009zm}.
As described above, the couplings in the theory are
consistent with internal momentum conservation at each vertex.
We have computed various mixed gauge-field and scalar four- and 
five-point amplitudes in the spontaneously-broken theory,  
and verified in all cases that the massive amplitudes are 
equivalent to the corresponding massless amplitudes 
with the replacement $\kk{a}{b} \rightarrow \kk{a}{b} - \kappa_a \kappa_b$,
where 
$\kappa_a = g( v_{i_a} - v_{i_{a+1}})$.
This is not initially obvious from the Feynman diagram calculation 
in the spontaneously-broken theory,
in which several different diagrams conspire to give the
appropriate expressions,
but is manifest from the dimensional reduction 
of the CHY representation of the amplitudes.
\para

What if the internal momenta $\kappa_a$ of the particles
span more than one dimension of the $M$-dimensional internal space?
In this case, the dimensionally-reduced amplitudes appear 
to correspond to massive amplitudes in a 
spontaneously-broken gauge theory in which several adjoint 
Higgs fields $H^I$,  $I=1, \cdots, h$  have nonzero 
vacuum expectation values
$\langle H^I \rangle = {\rm diag}(v^I_1, v^I_2, \cdots, v^I_N)$
that are mutually commuting.
In this theory, the off-diagonal gauge bosons $W_i^{~j}$ 
obtain masses  $g \left[ \sum_{I=1}^h \left(v^I_i-v^I_j\right)^2 \right]^{1/2}$.
We identify the internal momentum $\kappa^I_{ij}$ of $W_i^{~j}$ 
with $ g(v^I_i - v^I_j)$, automatically satisfying the 
triple vertex constraint $\kappa^I_{ij}+\kappa^I_{jk} = \kappa^I_{ik}$.
Although momentum conservation still requires $\sum_{a=1}^n \kappa^I_a = 0$, 
the constraint on the masses of the external particles is considerably
weakened compared to the case in which only one Higgs field gets a vev. 
\para

Finally, we recall that momentum conservation can be used to express
all kinematic invariants $\kk{a}{b}$ of an $n$-point amplitude
in terms of an independent set of $n(n-3)/2$ invariants
\be
k_1 \cdot k_c, \qquad k_2 \cdot k_c,  \qquad \qquad k_c \cdot k_d,\qquad\qquad
c, d \in \{3, \cdots, n-1\}.
\label{independent}
\ee
If we consider $n$-point amplitudes
with at most three massive particles ($a=1, 2$, and $n$),
then $\kappa_c = 0$ for $ c \in \{3, \cdots, n-1\}$
and all of the invariants in \eqn{independent} are unchanged under
$\kk{a}{b} \rightarrow \kk{a}{b} - \kappa_a \cdot \kappa_b$.
Consequently, the expressions for amplitudes with at most
three massive particles, when expressed in terms of 
$\epsk{a}{b}$, $\epseps{a}{b}$, and  the invariants (\ref{independent}), 
are identical to the corresponding massless amplitudes,
as found in ref.~\cite{Naculich:2015zha}.
For example, \eqns{fourpoint}{K}
reduce to the massless amplitude when $m_3$, and therefore $\kappa_3$, vanishes.
\para

In this note, we have argued that dimensionally-reduced 
tree-level gauge-theory amplitudes of massless particles possessing 
internal momentum $\kappa$ 
are equivalent to tree-level amplitudes of massive particles 
in a lower-dimensional gauge theory in which the gauge
symmetry is spontaneously broken by an adjoint Higgs field.
The same procedure can be applied to obtain amplitudes of
massive particles in other theories with a CHY representation, 
including gravity, Dirac--Born--Infeld theory, and the nonlinear $\sigma$ model.
It will be intriguing to determine the lower-dimensional
theories to which these massive amplitudes belong.
\para

\section*{Acknowledgments}
The author wishes to thank Freddy Cachazo, Song He, and Ellis Yuan for 
fruitful discussions.
This material is based upon work supported in part by the 
National Science Foundation under Grant No.~PHY14-16123.
\para

\end{document}